\documentclass[12pt]{article}
\usepackage{latexsym,amssymb,amstext,amsmath}
\usepackage{hyperref}
\usepackage{graphicx}
\usepackage{amsmath}
\usepackage{color}
\renewcommand{\Large}{\large} % smaller headlines

\allowdisplaybreaks
\begin{document}
%%%%%%%%%%%%%%%%%%%%%%%%%%%%%%%%%%%%%%%%%%%%%%%%%%%%%%%%%%%%%%%

%\begin{center}
%
%\textbf{Anisotropic expansion, dissipative hydrodynamics from kinetic theory}
%\end{center}
%\begin{center}
%\textbf{Priyanka Priyadarshini Pruseth
%\footnote{
%Present address: Department of Physics, Vesaj Patel College, 
%Sundargarh, India} and 
%Swapna Mahapatra}
%\vskip 6mm
%
%{\em Department of Physics, Utkal University,
%Bhubaneswar 751004, India.}\\
%
%{\tt pkpruseth@gmail.com}\;,\;\, {\tt swapna.mahapatra@gmail.com}
%\end{center}
%%%%%%%%%%%%%%%%%%%%%%%%%%%%%%%%%%%%%%%%%%%%%%%%%%%%%%%%%%%%%%%%%%
%%%%%%%%%%%%%%%%%%%%%%%%%%%%%%%%%%%%%%%%%%%%%%%%%%%%%%%%%%%%%
%
\begin{titlepage}
\bigskip
\begin{center}
%\vskip 6mm
%\vskip 3in
%\begin{document}
%\vskip 5pt
%\pagenumbering{roman}
\begin{center}
%{\textwidth 150mm \Large {\bf Anisotropic expansion, dissipative hydrodynamics 
%	from kinetic theory}}

{\Large \bf Anisotropic expansion, dissipative hydrodynamics from kinetic theory}
\\[10mm]
\end{center}
%\vskip 8mm

\textbf{Priyanka Priyadarshini Pruseth
\footnote{
Present address: Department of Physics, Vesaj Patel College,
Sundargarh, India} and
Swapna Mahapatra}

\vskip 6mm

{\em Department of Physics, Utkal University,
Bhubaneswar 751004, India.}\\

{\tt pkpruseth@gmail.com}\;,\;\, {\tt swapna.mahapatra@gmail.com}
\end{center}

\vskip .2in
%%%%%%%%%%%%%%%%%%%%%%%%%%%%%%%%%%%%%%%%%%%%%%%%%%%%%%%%%%%%%%%%
\begin{center}
%\begin{abstract}
{\bf Abstract }
\end{center}
\begin{quotation}
\noindent
%We discuss second and third-order hydrodynamic evolution equations for 
%shear stress tensor from kinetic theory.  
We consider Kasner space-time describing anisotropic three dimensional 
expansion of the fluid and obtain the dissipative evolution equations 
for shear stress tensor and energy density from kinetic theory. 
For this, we use the iterative solution of relativistic Boltzmann equation 
with relaxation time approximation. We show that our results 
for second and third order evolution equations 
reduce to those of one dimensional expansion case under suitable 
conditions for the anisotropic parameters in Kasner space-time.
%\end{abstract}

%%%%%%%%%%%%%%%%%%%%%%%%%%%%%%%%%%%%%%%%%%%%%%%%%%%%%%%%%%%%%%%%%%%%%%%
\end{quotation}
\vfill
%%%%%%%%%%%%%%%%%%%%%%%%%%%%%%%%%%%%%%%%%%%%%%%%%%%%%%%%%%%%
%\today
%%%%%%%%%%%%%%%%%%%%%%%%%%%%%%%%%%%%%%%%%%%%%%%%%%%%%%%%%%
\end{titlepage}
%%%%%%%%%%%%%%%%%%%%%%%%%%%%%%%%%%%%%%%%%%%%%%%%%%%%%

\section{Introduction}
\label{sec:introduction}
\setcounter{equation}{0}
%%%%%%%%%%%%%%%%%%%%%%%%%%%%%%%%%%%%%%%%%%%%%%%%%%%%%%%%%%

One of the fundamental question in physics is to
understand the properties of matter at extreme density and temperature in
the first few microseconds after the big bang. Such a state of matter is
known as Quark-Gluon-Plasma (QGP) state where the quarks and the gluons are
in deconfined state. A lot of progress has been made in undertanding the
properties and various aspects of the evolution of the strongly coupled QGP
through the heavy-ion collision experiments at RHIC 
\cite{Shuryak:2005, Kolb:2003KH} (see also \cite{Tannenbaum:2006}) as well
as in LHC (see \cite{Florkowski:2017} for a comprehensive review).
%Relativistic hydrodynamics is an effective theory describing the
%long-wavelength limit of the microscopic dynamics of a system. 
%%%%%%%%%%%%%%%%%%%%%%%%%%%%%%%%%%%%%%%%%%%%%%%%%%%%%%%%%%%%%%%%%%%%%%%
%The collective behavior of the hot and dense matter 
%(which is believed to have existed in the very early universe) created in 
%ultra-relativistic heavy-ion collisions has been studied quite extensively 
%within the framework of relativistic fluid dynamics. 
%%%%%%%%%%%%%%%%%%%%%%%%%%%%%%%%%%%%%%%%%%%%%%%%%%%%%%%%%%%%%%%%%%%%%%
Hydrodynamics plays an important role after the system undergoes a rapid 
thermalization and local thermal equilibrium is reached. Since it 
is difficult to solve the strongly coupled QCD, the qualitative 
features of the hydrodynamics regime in the evolution of QGP has been 
extensively studied by using AdS/CFT duality. This has provided 
an important framework for studying
the strongly coupled dynamics in a class of superconformal field theories,
in particular, ${\cal N} = 4$  super Yang-Mills theory
and the corresponding gravity dual description in AdS space-time
\cite{Malda:1998, WittenGubser:1998}. In the context of heavy ion
collisions, AdS/CFT correspondence has led to interesting results
like computation of shear viscosity of
finite temperature ${\cal N} = 4$ Supersymmetric Yang-Mills theory
\cite{Policastro:2001}, viscosity from gravity dual description involving
black holes in AdS space  \cite{Kovtun:2005} etc.
%%%%%%%%%%%%%%%%%%%%%%%%%%%%%%%%%%%%%%%%%%%%%%%%%%%%%%%%%%%%%%%%%%%%%%%%
As we know, perfect local equilibrium system is described by ideal
fluid dynamics. For small departures from equilibrium, the system is
described by dissipative fluid dynamics. The nonrelativistic 
limit of hydrodynamics can be described by 
Navier-Stokes equations involving the energy density, pressure of the fluid 
as well as the shear and bulk viscosities. These equations are relevant 
for describing the fluid at low energy.
%and the hydrodynamics description can be understood in terms of the 
%mean free path $l_{mfp}$ which is basically the underlying microscopic scale.  
However, in the context of 
ultrarelativistic heavy-ion collisions at RHIC and LHC, one needs 
to use relativistic hydrodynamics to study the evolution and properties 
of QGP describing fluid at high energy. 
%Apart from heavy-ion collisions, 
%relativistic hydrodynamics also finds its applicability in various 
%discussions in the areas of astrophysics and cosmology.

The microscopic dynamics describing 
processes characterized by $kl_{mfp}$ ($k$ being the momentum scale 
and $l_{mfp}$ is the mean free path of the system) has been studied 
in great detail using the relativistic hydrodynamics setup. 
In the framework of gradient expansion in relativistic dissipative 
hydrodynamics, 
the zeroth order theory is described by ideal hydrodynamics. 
First order theory is described by relativisitc Navier-Stokes 
equations. In first order hydrodynamics, due to the 
lack of an initial value formulation, 
signals can be transmitted with arbitrarily high speed thereby violating 
causality. The system in this case is described by parabolic equations. 
The first order theory has
been extended by M\"uller \cite{Muller:67} and independently 
by Israel and Stewart(IS)
\cite{Israel:197679IS} by including the second order gradient terms thereby
preserving causality in the resulting relativistic hydrodynamics equations.
The set of transport coefficients are extended in the second order
hydrodynamics and the resulting equations become hyperbolic. However, the  
M\"uller-Israel-Stewart theory does not contain all the corrections 
to second order in gradient. Subsequently, 
the authors in refs 
\cite{Baier:2007BRSSS, Sayantani:2007BHMR} obtained additional terms in 
the stress-energy tensor by 
utilising the conformal symmetries of the theory and the new transport 
coefficients were explicitly determined. The relaxation time 
(a transport coefficient 
in second order viscous hydrodynamics) has also been computed from the 
analysis of the regularity of the dual geometry \cite{Heller:2007HJ}. 

In the second order Israel-Stewart (IS) theory, the equations in relativistic 
dissipative fluid dynamics have been obtained using the second 
moment of Boltzmann equation in the kinetic theory description and 
Grad's 14-moment expansion \cite{Grad} for the phase-space distribution 
function. 
However, 14-moment approximation used in IS theory does not provide
a unique theory as it leads to multiple fluid dynamical equations
with similar general structure but different transport coefficients
\cite{Denicol:2010xn, Denicol:2012}. For one dimensional Bjorken 
expansion \cite{Bjorken:1983}, it has been discussed  
that the second order 
IS theory has also some unphysical features like reheating of the 
expanding medium \cite{Muronga:2003ta}, negative pressure 
\cite{Martinez:2009mf} for large viscosity and small initial expansion time; 
large ratio of shear viscosity to entropy 
density 
\cite{El:2008yy} etc.
In order to address these problems with IS theory, higher order 
corrections in the dissipative hydrodynamics were considered to study 
their effects \cite{El:2009vj}.
Though the resulting evolution equation 
did not contain all possible terms in the second order dissipative 
hydrodynamics, the solutions with higher order correction indicated better
agreement with the results from kinetic transport theory.   
%%%%%%%%%%%%%%%%%%%%%%%%%%%%%%%%%%%%%%%%%%%%%%%%%%%%%%%%%%%%%%%%%%%%

Second-order dissipative equations have been 
derived from relativistic Boltzmann equation (BE)  
using gradient expansion of the distribution function thereby including
nonlocal effects in the collision term \cite{Jaiswal:2012qm}. In the above 
setup, while deriving the evolution equations, the authors use the 
definition of dissipative current directly instead of 
using the second moment of BE and it is important 
to note that the evolution equations included all possible second 
order terms allowed by the symmetry. Subsequently considering the nonlocal 
collision term, the relativistic 
third order dissipative evolution equation for the shear stress tensor has 
been derived from kinetic theory without using Grad's 
14-moment approximation and second moment of BE \cite{AJ}. 
In the above formalism, BE has been solved 
iteratively in the relaxation 
time approximation (RTA) for the collision term 
and the nonequilibrium phase space distribution function has been obtained
(see also \cite{Jaiswal:2013npa} for the discussion in
the context of first order and second order evolution equations).
The formulation of RTA for the collision term was proposed in a very 
interesting paper by Anderson and Witting \cite{Anderson-Witting} 
for solving the relativistic BE as a power series in the relaxation 
time and the transport coefficients 
for a single component gas were obtained and compared with the relativistic 
Grad's moment method. 
Iterative solution method has also been useful in obtaining 
higher order corrections to entropy four current \cite{CJPR:2014}.
Hydrodynamics gradient expansion in higher orders has been discussed 
in ref.\cite{Grozdanov:2015}, where, third order corrections in conformal as 
well as nonconformal hydrodynamics of neutral fluids have been investigated. 
In the case of nonrelativistic systems, 
higher order constitutive equations from kinetic theory have been discussed 
earlier in ref.\cite{Cha}. 
%%%%%%%%%%%%%%%%%%%%%%%%%%%%%%%%%%%%%%%%%%
There has been progress in hydrodynamical 
formulations from various other approaches 
\cite{Manes:2022, Cantarutti:2020, Jaiswal-rev:2021}.
%%%%%%%%%%%%%%%%%%%%%%%%%%%%%%%%%%%%%%%%%%%%%%%%%%%%%%%%%%%%%%%%%%%%
In this work, we consider a generalization of Bjorken's one dimensional 
expansion to three dimensional anisotropic expansion of the relativistic 
fluid, where the local rest frame (LRF) of the 
anisotropically expanding fluid is described by time 
dependent Kasner space-time. 
For time dependent AdS/CFT correspondence, Kasner 
space-time has been studied earlier in the 
context of anisotropic expansion of the RHIC and LHC fireball, where 
explicit expressions for the hydrodynamic quantities have been obtained 
in first and second order relativistic viscous hydrodynamics 
and gravity dual description has been studied 
\cite{Nakamura:2006SNK, ppsm:2020}. Collisionless Boltzmann 
equation in Kasner space-time and its relation to 
anisotropic hydrodynamics has been discussed in ref.\cite{Jaiswal:2017}.
Though Kasner space-time is a curved space-time, Sin {\it et al} 
\cite{Nakamura:2006SNK} have shown that under a well controlled 
approximation, it can be considered as the LRF of the anisotropically 
expanding fluid on Minkowski space-time. 
Here, we study the relativistic BE in RTA for the collision term. 
We extend the results of 
ref. \cite{AJ} to three dimensional expansion case. Using the 
iterative solution of BE for the nonequilibrium distribution function,
we obtain the dissipative evolution equations for the shear stress tensor 
and energy density to second and third order in gradients from kinetic 
theory in Kasner space-time. These expressions have not 
been obtained before. We show that our results reduce to that of one
dimensional expansion case under suitable conditions for the Kasner 
parameters. 

The paper is organized as follows: Section 1 contains the introduction 
and motivation for the present study. Section 2 deals with the basic 
formalism for solving the relativistic BE 
iteratively in RTA for the collision term. 
We discuss the basic set-up in Minkowski space-time.  
In section 3, we generalize the 
one dimensional Bjorken expansion to three dimensional anisotropic 
expansion by considering Kasner space-time. Using the iterative 
solutions of the Boltzmann equation for the nonequilibrium distribution
function, we obtain the second and third order evolution equations 
for the shear stress tensor ind energy density in terms of Kasner 
parameters. We also show that our evolution equations  
agree with the one dimensional Bjorken expansion case 
in the appropriate limit of the Kasner parameters. We summarise 
and discuss the future perspective in section 4. 
%%%%%%%%%%%%%%%%%%%%%%%%%%%%%%%%%%%%%%%%%%%%%%%%%%%%%%%%%%%%%%%%%%%%%%%

\section{Kinetic theory and iterative solution of relativistic 
Boltzmann equation}
\label{sec:Iterative solution}
\setcounter{equation}{0}

As we know, the macroscopic state of a system in relativistics 
fluid dynamics is described by the energy-momentum tensor and 
the corresponding conservation law plays an important role in 
the hydrodynamic evolution of a system. The conserved energy-momentum 
tensor $T^{\mu\nu}$ can be decomposed as, 
\begin{eqnarray}
\label{NTD}
T^{\mu\nu} = \epsilon u^{\mu} u^{\nu} - P \Delta^{\mu\nu} + \pi^{\mu\nu},
\end{eqnarray}
where, $\epsilon$, $P$ and $\pi^{\mu\nu}$ represent the energy density, 
pressure and shear stress tensor respectively. The bulk viscosity 
vanishes for a system of massless particles and the corresponding 
theory is conformal. We work in the Landau frame.  
The projection operator $\Delta^{\mu\nu}$ defined on the space 
orthogonal to the fluid velocity is given by: 
$\Delta^{\mu\nu} = g^{\mu\nu}-u^{\mu} u^{\nu}$, where the metric 
tensor $g^{\mu\nu} = diag (+, -, -, -)$ and $u^{\mu}$ is the 
fluid 4-velocity. $u^{\mu}$ is an eigen vector of the energy-momentum 
tensor ($T^{\mu\nu} u_{\nu} = \epsilon u^{\mu}$) and it satisfies the 
following properties: $\Delta^{\mu\nu} u_{\mu} = 0 = \Delta^{\mu\nu} u_{\nu}; 
\,\, \Delta^{\mu\nu} \Delta^{\alpha}_{\nu} = \Delta^{\mu\alpha}$. 
In the local rest frame of the fluid, the 4-velocity is given by 
$u^{\mu} = (1, 0, 0, 0)$. 
Projecting the energy-momentum tensor conservation equation  
%$\partial_\mu T^{\mu\nu} =0$ 
in the direction parallel and perpendicular 
to the fluid 4-velocity results in the following fundamental 
equations for a relativistic viscous fluid: 

\begin{eqnarray}\label{evol}
\dot\epsilon + (\epsilon+P)\theta - \pi^{\mu\nu} \sigma_{\mu\nu} &=0,\nonumber\\
(\epsilon+P)\dot u^\alpha - \nabla^\alpha P + \Delta^\alpha_\mu \partial_\nu
\pi^{\mu\nu} &= 0.
\end{eqnarray}
We use the notations of ref.\cite{AJ}. Here
$\dot \epsilon = u^\mu\partial_\mu \epsilon $ denotes the comoving
derivative of the energy density, $\theta 
\equiv \nabla_\mu u^\mu$ is the expansion
scalar, $\nabla^\alpha\equiv \Delta^{\mu\alpha}\partial_\mu$ denotes 
the space-like derivative and $\sigma^{\mu\nu} \equiv 
\Delta^{\langle\mu}u^{\nu\rangle} = 
\Delta^{\mu\nu}_{\alpha\beta} \nabla^{\alpha} u^{\beta}$. 
Symmetrization is defined by $A^{(\mu}B^{\nu )}
\equiv \frac{1}{2} (A^\mu B^\nu + A^\nu B^\mu)$. 
%$\partial_{\mu} = u^{\mu} D + \nabla_{\mu} $ and 
%$D \equiv u^{\mu} \partial_{\mu}$.  

%%%%%%%%%%%%%%%%%%%%%%%%%%%%%%%%%%%%%%%%%%%%%%%%%%%%%%%%%%%
%decomposed inexpressed in terms of single 
%particle, phase space distribution function and tensor decomposed into 
%hydrodynamic variables \cite{deGroot}. 
%The energy-momentum tensor is given by,
%\begin{align}\label{NTD}
%T^{\mu\nu} &= \!\int\! dp \ p^\mu p^\nu\, f(x,p) = \epsilon u^\mu u^\nu 
%- P\Delta ^{\mu \nu} + \pi^{\mu\nu},
%\end{align}
%where $dp\equiv g d{\bf p}/[(2 \pi)^3|\bf p|]$, $g$ represents the 
%degeneracy factor, $p^\mu$ is the particle four-momentum and $f(x,p)$ is 
%the phase-space distribution function. 
%%%%%%%%%%%%%%%%%%%%%%%%%%%%%%%%%%%%%%%%%%%%%%%%%%%%%%%%%%%%%%%%

Relativistic hydrodynamics can be derived from kinetic theory where 
the fundamental equation is Boltzmann equation. 
Relativistic BE in kinetic theory is given by
\cite{deGroot}, 
\begin{equation}\label{BE}
p^{\mu} \partial_{\mu} f(x, p) =  {\cal C}[f](x, p)
\end{equation}
where $p^{\mu}$ is the particle 4-momentum, $p^{\mu} = 
(p^0, {\bf p}$) with $p^0 = {\sqrt {{\bf p}^2 + m^2}}$, $f(x, p)$ is the 
one particle phase space distribution function and 
${\cal C}[f](x, p)$ is the collision term. 
Right hand side of the 
above equation becomes zero if the collision between particles is neglected. 
In the kinetic theory, the energy-momentum tensor is expressed in terms
of the distribution function $f(x, p)$ and particle 4-momentum $p^{\mu}$:

\begin{eqnarray}\label{NTD}
T^{\mu\nu} &= \!\int\! dp \ p^\mu p^\nu\, f(x,p)
\end{eqnarray}
where $dp\equiv g d{\bf p}/[(2 \pi)^3|\bf p|]$ and g is the number of
internal degrees of freedom and we are considering a system of massless
particles.

In the RTA, the collision term is given by
\cite{Anderson-Witting}, 
\begin{equation}\label{collision}
{\cal C}[f](x, p) = \frac{u \cdot p}{\tau_R} (f(x,p) - f_{eq}(x,p))
\end{equation}
where, $u\cdot p = u_{\mu} p^{\mu}$, $\tau_R$ is the relaxation time, 
$f_{eq}(x, p)$ is the equilibrium distribution function and 
the deviation $\delta f$ from the equilibrium value is assumed to be 
small. We write,
\begin{equation}\label{deviation} 
f(x, p) = f_{eq} (x, p) +  \delta f(x, p)
\end{equation}

The shear stress tensor $\pi^{\mu\nu}$ in the decomposition of $T^{\mu\nu}$ 
can be calculated in terms of $\delta f$ which is the deviation from 
the equilibrium distribution function. 
$\pi^{\mu\nu}$ can be written as, 
\begin{eqnarray}\label{FSE}
\pi^{\mu\nu} &= \Delta^{\mu\nu}_{\alpha\beta} \int dp \, p^\alpha p^\beta\, 
\delta f,
\end{eqnarray}

where,
\begin{equation}\label{sym-proj-op}
\Delta^{\mu\nu}_{\alpha\beta}\equiv 
\frac{1}{2}(\Delta^{\mu}_{\alpha}\Delta^{\nu}_{\beta} + 
	\Delta^{\mu}_{\beta}\Delta^{\nu}_{\alpha} - 
	\frac{2}{3}\Delta^{\mu\nu}\Delta_{\alpha\beta}) 
\end{equation}
is a traceless symmetric projection operator orthogonal to 
$u^\mu$ and it satisfies the following properties: 
$\Delta^{\mu\nu}_{\alpha\beta} \Delta_{\mu\nu} = 0 = 
\Delta^{\mu\nu}_{\alpha\beta} \Delta^{\alpha\beta}$, 
$\Delta^{\mu\nu}_{\rho\sigma}\Delta^{\rho\sigma}_{\alpha\beta} 
= \Delta^{\mu\nu}_{\alpha\beta}$.

In order to obtain the expression for the nonequilibrium part  $\delta f$ 
appearing above in the expression for $\pi^{\mu\nu}$, one solves 
the BE iteratively in RTA \cite{Jaiswal:2013npa}.  
In this formalism, one writes the deviation $\delta f$ in a gradient 
expansion \cite{Chapman}:
$\delta f = \delta f^{(1)} + \delta f^{(2)} + \delta f^{(3)} + \cdots $,
where $\delta f^{(1)}$ is first order in gradient, $\delta f^{(2)}$, 
$\delta f^{(3)}$ are second and third order in gradients respectively. 

Using the expression for the collision functional in the RTA 
\cite{Anderson-Witting},
\begin{eqnarray}\label{collision-RTA}
{\cal C}[f] = -u\!\cdot\! p\frac{\delta f}{\tau_R},
\end{eqnarray}
the relativistic BE has been solved iteratively 
\cite{Jaiswal:2013npa}, where,   
\begin{eqnarray}\label{F1F2}
f_1 &= f_{eq} -\frac{\tau_R}{u\!\cdot\! p} \, p^\mu \partial_\mu f_0, \quad 
\nonumber \\
f_2 &= f_{eq} -\frac{\tau_R}{u\!\cdot\! p} \, p^\mu \partial_\mu f_1, 
~~\, \cdots
\end{eqnarray}
with the notation $f_n = f_{eq} + \delta f^{(1)} + 
\delta f^{(2)} + \cdot + \delta f^{(n)}$.
From here, one obtains the expressions for the deviation $\delta f$ 
in an gradient expansion as \cite{AJ}: 
%To first and second-order in derivatives \cite{AJ}, one obtain
\begin{eqnarray}
\delta f^{(1)} &= -\frac{\tau_R}{u\!\cdot\! p} \, p^\mu \partial_\mu f_{eq} 
\label{FOC} \\
\delta f^{(2)} &= \frac{\tau_R}{u\!\cdot\! p}p^\mu p^\nu
\partial_\mu\Big(\frac{\tau_R}{u\!\cdot\! p} 
\partial_\nu f_{eq}\Big). \label{SOC}
\end{eqnarray}
The shear stress tensor to first order in gradients can be calculated 
by using the expression for $\delta f^{(1)}$ given above 
[eqn.(\ref{FOC})] in eqn.(\ref{FSE}) resulting  
$\pi^{\mu\nu} = 2\tau_R\beta_\pi\sigma^{\mu\nu}, 
\quad\beta_\pi = \frac{4}{5}P$.  
Using the above results, one can obtain the expressions 
for the shear stress tensor and its evolution to higher orders 
in derivatives.   

The above hydrodynamics set up has been dicussed in 
Minkowski space-time.
In the next section, we shall discuss the anisotropic expansion 
of the fluid, in which, Kasner space-time has been considered as the 
local rest frame of the fluid.  
It is important to note that though Kasner space-time is a curved 
space-time, Sin {\it et al} \cite{Nakamura:2006SNK} have shown that 
under a well controlled
approximation, it can be considered as the LRF of the anisotropically
expanding fluid on Minkowski space-time. The classical Boltzmann equation in
curved space-time involves the Christoffel symbol, which is given by,
\begin{equation}
	(p^{\mu} \partial_{\mu}-
\Gamma^{\lambda}_{\mu \nu} 
	p^{\mu} p^{\nu} \partial_{\lambda}) f(p,x)= {\mathcal C} [f]
\end{equation}
The stress energy tensor is defined as, 
\begin{equation}
T^{\mu\nu} = \int {\sqrt{-g}}\frac{d^3p}{p^0} p^{\mu}p^{\nu} f(x, p)
\end{equation}
where, $g$ is the determinant of the metric tensor $g_{\mu\nu}$. 
One obtains the hydrodynamic equations by taking the moments {\it w.r.t.}
to the particle momentum.  
Baier {\it et al} have discussed the set up in curved background in 
ref.\cite{Baier:2007BRSSS}. Hence instead of repeating, 
We refer to the readers ref. \cite{Baier:2007BRSSS} 
for the discussion in a general curved background.
%%%%%%%%%%%%%%%%%%%%%%%%%%%%%%%%%%%%%%%%%%%%%%%%%
In the present framework, the effect of a general background 
space-time 
can be accounted for by replacing the partial derivative 
$\partial_\mu$ with the covariant derivative $D_\mu$ throughout. 
For example, we have, 
$\dot u^{\alpha} = u^{\mu} D_{\mu} u^{\alpha}$ and  
the covariant derivative of a vector is given by ,
\begin{eqnarray}\label{covariant-d}
D_{\mu} A^{\nu} = \partial_\mu A^{\nu} + \Gamma^{\nu}_{\mu\rho} A^{\rho}; 
\,\,\,
D_{\mu} A_{\nu} = \partial_\mu A_{\nu} - \Gamma^{\rho}_{\mu\nu} A_{\rho}
\end{eqnarray} 
In the next section, we use the covariant derivative for 
computing various expressions where 
we consider Kasner space-time as the local rest frame 
(LRF) of the fluid.
%%%%%%%%%%%%%%%%%%%%%%%%%%%%%%%%%%%%%%%%%%%%%%%%%%%%%%
%%%%%%%%%%%%%%%%%%%%%%%%%%%%%%%%%%%%%%%%%%%%%%%%%%%%%%%%%%%%%%%%%%%%%%%
\section{Evolution equations for shear stress tensor and Kasner space-time}
%%%%%%%%%%%%%%%%%%%%%%%%%%%%%%%%%%%%%%%%%%%%%%%%%%%%%%%%%%%%%%%%%%%

In order to obtain the evolution equation for the shear stress tensor
to higher orders, 
one needs to compute the comoving derivative of $\delta f$ 
(which is the deviation from the equilibrium distribution function) 
expanded in powers of space-time derivatives. In particular, one has,   
%\cite {Jaiswal:2013npa}:
\begin{equation}
\dot\pi^{\langle\mu\nu\rangle} = \Delta^{\mu\nu}_{\alpha\beta} 
\int dp\, p^\alpha p^\beta\, \delta\dot f, \label{SSE}
\end{equation}
where we have used the standard notation $A^{\langle\mu\nu\rangle}\equiv 
\Delta^{\mu\nu}_{\alpha\beta}A^{\alpha\beta}$. To second order in 
gradients, the evolution equation for the shear tensor is given by 
\cite{Jaiswal:2013npa}, 
\begin{equation}\label{SOSHEAR}
\dot{\pi}^{\langle\mu\nu\rangle} \!+ \frac{\pi^{\mu\nu}}{\tau_\pi}\!= 
2\beta_{\pi}\sigma^{\mu\nu}
\!+2\pi_\gamma^{\langle\mu}\omega^{\nu\rangle\gamma}
\!-\frac{10}{7}\pi_\gamma^{\langle\mu}\sigma^{\nu\rangle\gamma} 
\!-\frac{4}{3}\pi^{\mu\nu}\theta,
\end{equation}
where $\omega^{\mu\nu}\equiv \frac{1}{2}(\nabla^\mu u^\nu-\nabla^\nu u^\mu)$
is the fluid vorticity.
In the derivation of this equation, one uses the expression for 
$\delta f^{(1)}$ in the expansion of the particle distribution 
function $f$ and keeps terms upto quadratic in the gradient 
expansion. In the above, the Boltzmann relaxation time $\tau_R$ has 
been replaced by the shear relaxation time $\tau_{\pi}$ which is 
a second order transport coefficient and can be expressed as 
$\tau_{\pi} = \frac{\eta}{\beta_{\pi}}$ where $\eta$ is the first order 
transport coefficient and $\beta_{\pi}$ is related to the pressure 
$P$ of the fluid. 

Similarly, using the expressions for $\delta f^{(1)}$ (eqn.\ref{FOC}), 
$\delta f^{(2)}$ (eqn.\ref{SOC}), computing their comoving derivatives 
and keeping terms upto cubic order in derivatives, the 
third order evolution equation for $\pi^{\mu\nu}$ has 
been obtained by Jaiswal \cite{AJ}:

\begin{eqnarray}\label{TOSHEAR}
\dot{\pi}^{\langle\mu\nu\rangle} =& -\frac{\pi^{\mu\nu}}{\tau_\pi}
+2\beta_\pi\sigma^{\mu\nu}
+2\pi_{\gamma}^{\langle\mu}\omega^{\nu\rangle\gamma}
-\frac{10}{7}\pi_\gamma^{\langle\mu}\sigma^{\nu\rangle\gamma}  
-\frac{4}{3}\pi^{\mu\nu}\theta
+\frac{25}{7\beta_\pi}\pi^{\rho\langle\mu}\omega^{\nu\rangle\gamma}\pi_{\rho\gamma}
-\frac{1}{3\beta_\pi}\pi_\gamma^{\langle\mu}\pi^{\nu\rangle\gamma}\theta \nonumber \\
&-\frac{38}{245\beta_\pi}\pi^{\mu\nu}\pi^{\rho\gamma}\sigma_{\rho\gamma}
-\frac{22}{49\beta_\pi}\pi^{\rho\langle\mu}\pi^{\nu\rangle\gamma}\sigma_{\rho\gamma} 
-\frac{24}{35}\nabla^{\langle\mu}\left(\pi^{\nu\rangle\gamma}\dot u_\gamma\tau_\pi\right)
+\frac{4}{35}\nabla^{\langle\mu}\left(\tau_\pi\nabla_\gamma\pi^{\nu\rangle\gamma}\right) \nonumber \\
&-\frac{2}{7}\nabla_{\gamma}\left(\tau_\pi\nabla^{\langle\mu}\pi^{\nu\rangle\gamma}\right)
+\frac{12}{7}\nabla_{\gamma}\left(\tau_\pi\dot u^{\langle\mu}\pi^{\nu\rangle\gamma}\right) 
-\frac{1}{7}\nabla_{\gamma}\left(\tau_\pi\nabla^{\gamma}\pi^{\langle\mu\nu\rangle}\right)
+\frac{6}{7}\nabla_{\gamma}\left(\tau_\pi\dot u^{\gamma}\pi^{\langle\mu\nu\rangle}\right) \nonumber \\
&-\frac{2}{7}\tau_\pi\omega^{\rho\langle\mu}\omega^{\nu\rangle\gamma}\pi_{\rho\gamma}
-\frac{2}{7}\tau_\pi\pi^{\rho\langle\mu}\omega^{\nu\rangle\gamma}\omega_{\rho\gamma} 
-\frac{10}{63}\tau_\pi\pi^{\mu\nu}\theta^2
+\frac{26}{21}\tau_\pi\pi_\gamma^{\langle\mu}\omega^{\nu\rangle\gamma}\theta.
\end{eqnarray}

%The above third order equation  
%contains three second order and fourteen third order terms. 
%%%%%%%%%%%%%%%%%%%%%%%%%%%%%%%%%%%%%%%%%%%%%%%
Note that the above equation was formulated within the 
framework of kinetic theory for a system of massless particles which has 
conformal symmetry. For such a system, dissipation due to bulk viscosity 
and heat current can be neglected.
%%%%%%%%%%%%%%%%%%%%%%%%%%%%%%%%%%%%%%%%%%%%%%%%%
Using entropy current and 
second law of thermodynamics, El {\it etal} (ref.\cite{El:2009vj})
have obtained the evolution equation before. However, in the context 
of one dimensional Bjorken expansion, the evolution equation 
obtained there misses out many more terms. 
%%%%%%%%%%%%%%%%%%%%%%%%%%%%%%%%%%%%%%%%%%%%%%%%%%%%%%%%%%%%%%%%%%%%%%%%%
Next, we study the evolution equation to second 
and third order for the shear stress tensor and energy density 
in the context of three dimensional anisotropic
expansion of the conformal fluid. 
%%%%%%%%%%%%%%%%%%%%%%%%%%%%%%
We consider the generalisation of Bjorken's 1-dimensional expansion to 
3-dimensional expansion of the fluid in order to 
connect to the realistic description of the RHIC and LHC fireball and  
for this, we consider Kasner space-time as the LRF of the fluid
\cite{Nakamura:2006SNK, ppsm:2020, Jaiswal:2017}.
The metric is \cite{Kasner:1921}: 
\begin{eqnarray}\label{Kasner}
ds^{2} = (d \tau)^{2} - \tau^{2a}(dx_{1})^{2} - \tau^{2b}(dx_{2})^{2}
- \tau^{2c}(dx_{3})^{2}
\end{eqnarray}

Here $x_1, x_2, x_3$ are the comoving coordinates, $\tau$ is 
the proper time and $a, b, c$ are constants known as Kasner parameters. 
The Kasner parameters satisfy
the conditions,
\begin{eqnarray}\label{Kasner-condn}
a + b + c = 1, \hspace{20pt}a^2 + b^2 + c^2 = 1 
\end{eqnarray}
The above metric is an exact solution of vacuum Einstein's
equation and it describes a homogeneous and anisotropic
expansion of the Universe. The physical quantities are assumed to 
depend only on proper time $\tau$.
The nonzero components of the affine connection 
for the Kasner metric are given by, 
\begin{eqnarray}\label{affine}
\Gamma^{\tau}_{x_1 x_1} = a \tau^{2 a -1},\, \Gamma^{\tau}_{x_2 x_2}
= b \tau^{2 b -1},\, \Gamma^{\tau}_{x_3 x_3} = c \tau^{2 c -1},\nonumber \\ 
\Gamma^{x_1}_{x_1 \tau} = \frac{a}{\tau}, \,
\Gamma^{x_2}_{x_2 \tau} = \frac{b}{\tau}, \,\Gamma^{x_3}_{x_3 \tau} = 
\frac{c}{\tau}
\end{eqnarray}

The Ricci tensor turns out
to be zero upon using Kasner conditions $a + b + c =1$; $a^2 + b^2 + c^2 =1$.
Ricci tensor for Kasner space-time is give by,
\begin{eqnarray}
R_{00} & = & \frac{1}{\tau^2} [ (a + b + c) - (a^2 + b^2 + c^2)] \\ \nonumber
R_{11} & = & a [(a + b + c) - 1] \tau^{2a-2} \\ \nonumber
R_{22} & = & b [(a + b + c) - 1] \tau^{2b-2} \\ \nonumber
R_{33} & = & c [(a + b + c) - 1] \tau^{2c-2} 
\end{eqnarray}
where, $1, 2, 3$ corresponds to $x_1, x_2, x_3$ coordinates, and
$0$ corresponds to $\tau$ coordinate. As one can see, the components
vanish upon using Kasner conditions. The nonzero components of Riemann tensor
are given by,
\begin{eqnarray}        
R_{0101} &=& (1-a)a \tau^{2a-2}, \, R_{0202} = (1-b) b\tau^{2b-2} 
        \\ \nonumber 
        R_{0303} &=& (1-c) c \tau^{2c-2}, \,
        R_{1212} = a b \tau^{2a+2b-2} \\ \nonumber
        R_{1313} &=& a c \tau^{2a + 2c-2} , \, R_{2323} = 
        b c \tau^{2b + 2c -2} 
\end{eqnarray}
These expressions become zero for the Bjorken case corresponding to
$a=1, b=0, c=0$. So the terms involving $R^{\mu\nu\rho\sigma}$
could contribute to the shear tensor $\pi^{\mu\nu}$ with a coefficient
$\kappa$ (we refer to eqn. 3.12 in ref. \cite{Baier:2007BRSSS}).
However, the general expression for the shear stress tensor $\pi^{\mu\nu}$
when derived from kinetic theory does not contain the $\kappa$ term which
would have involved the Riemann tensor (we refer to eqn. 5.23 of
the above reference \cite {Baier:2007BRSSS} and subsequent discussion). 
Boltzmann equation does not contain the term involving $\kappa$. 
Hence, there is no inconsistency in the order of the derivative expansion.

We compute various quantities appearing in the second and third order 
evolution equations for the shear stress tensor. 
%%%%%%%%%%%%%%%%%%%%%%%%%%%%%%%%%%%%%%%%%%%%%%%%%
As mentioned in the previous section, here
we use the notations involving covariant 
derivative $D_{\mu}$ (as defined in eqn. (\ref {covariant-d})) instead 
of partial 
derivative $\partial_{\mu}$.
We have, 
\begin{eqnarray}
\nabla_{\mu}u_{\nu} = \Delta^{\rho}_{\mu} D_{\rho}u_{\nu},
\nonumber \\
\sigma^{\mu\nu}  = \Delta^{\mu\nu}_{\alpha\beta} \nabla^{\alpha} u^{\beta}=
\Delta^{\mu\nu}_{\alpha\beta} \Delta^{\alpha\rho} D_{\rho}u^{\beta}
\end{eqnarray}

%%%%%%%%%%%%%%%%%%%%%%%%%%%%%%%%%%%%%%%%%%%%%
We obtain the 
components of the projection operator as,
\begin{equation}\label{projector}	
\Delta ^{\mu \nu}\equiv\mathrm{diag}(0, 
	- \frac{1}{\tau^{2a}},
	- \frac{1}{\tau^{2b}}, - \frac{1}{\tau^{2c}})
\end{equation}
%%%%%%%%%%%%%%%%%%%%%%%%%%%%%%%%%%%%%%%%%%%%%%%%%%%%%%%%
Components of the shear tensor are obtained as,  
%%%%%%%%%%%%%%%%%%%%%%%%%%%%%%%%%%%%%%%%%%%%%%%%%%%%%%%%%%%%%

\begin{eqnarray}\label{secondterm}
\sigma^{\tau\tau} = 0,\,
\sigma^{x_1 x_1} = \frac{-2a+b+c}{3}\tau^{-2a-1},\,\nonumber\\
\sigma^{x_2 x_2} = \frac{a-2b+c}{3}\tau^{-2b-1},\,\nonumber\\
\sigma^{x_3 x_3} = \frac{a+b-2c}{3}\tau^{-2c-1}
\end{eqnarray}
%%%%%%%%%%%%%%%%%%%%%%%%%%%%%%%%%%%%%%%%%%%%%%%%%%%%%%%%%%%%%%%%%
where we have used Kasner conditions and,
%%%%%%%%%%%%%%%%%%%%%%%%%%%%%%%%%%%%%%%%%%%%%%%%%%%%%%%%%%%%%%

\begin{equation}\label{theta}
\theta\equiv \nabla_\mu u^\mu = \Delta^{\nu}_{\mu}D_{\nu}u^{\mu} 
=\frac{a+b+c}{\tau}.
\end{equation}

%%%%%%%%%%%%%%%%%%%%%%%%%%%%%%%%%%%%%%%%%%%%%%%%%%%%%%%%%%%%
Upon using Kasner condition, $\theta$ becomes $\frac{1}{\tau}$ 
which is the same as in one dimensional Bjorken expansion case. 
The shear stress tensor is diagonal. 
%%%%%%%%%%%%%%%%%%%%%%%%%%%%%%%%%%%%%%%%%
We assume that it can be 
characterized by a function $\pi$ and the  Kasner parameters in the 
following form:   
$\pi^{\mu\nu}\equiv\mathrm{diag}(0,-\pi\tau^{-2a},
\frac{\pi}{2}\tau^{-2b},\frac{\pi}{2}\tau^{-2c})$. One can check 
that $\pi^{\mu\nu}$ is traceless 
%%%%%%%%%%%%%%%%%%%%%%%%%%%%%%%%%%
as we are considering a conformal fluid. 
%%%%%%%%%%%%%%%%%%%%%%%%%%%%%%%%%%%%%%%%%%%%

The components of $\dot{\pi}^{\langle \mu\nu \rangle}$ are given by, 
\begin{eqnarray}{\label{zeroterm}}
\dot{\pi}^{\langle \tau\tau \rangle}= 0, \,
\dot{\pi}^{\langle x_1x_1 \rangle}=\frac{-1}{\tau^{2a}}\frac{d\pi}{d\tau}, \,\nonumber\\
\dot{\pi}^{\langle x_2x_2 \rangle}=\frac{1}{2\tau^{2b}}\frac{d\pi}{d\tau}, \,
\dot{\pi}^{\langle x_3x_3 \rangle}=\frac{1}{2\tau^{2c}}\frac{d\pi}{d\tau}
\end{eqnarray}

Components of other terms appearing in the second order 
evolution equation are obtained as,
\begin{eqnarray}{\label{fourthterm}}
\pi_\gamma^{\langle \tau}\sigma^{\tau\rangle\gamma}=0,
\pi_\gamma^{\langle x_1}\sigma^{x_1\rangle\gamma}=\frac{\pi(b+c-2a)}
{6 \tau^{2a+1}},\nonumber\\
\pi_\gamma^{\langle x_2}\sigma^{x_2 \rangle\gamma}=
\frac{\pi(a+b-2c)}{6 \tau^{2b+1}},\nonumber\\
\pi_\gamma^{\langle x_3}\sigma^{x_3 \rangle\gamma}=\frac{\pi(a+c-2b)}
{6 \tau^{2c+1}}
\end{eqnarray}
Substituing the above expressions, the second order evolution equations
are given by
\begin{eqnarray}
\frac{d\epsilon}{d\tau} &= -\frac{1}{\tau}\left(\epsilon + P \right)  
-\frac{\pi(b+c-2a)}{2\tau} \label{BED1}, \\
\frac{d\pi}{d\tau} &= - \frac{\pi}{\tau_\pi} - 
\frac{2\beta_\pi (-2a+b+c)}{3\tau} - \frac{\pi(38a+23b+23c)}{21\tau}, 
\label{Bshear1}
\end{eqnarray}
Note that, adding the three equations for the nonzero components of
$\dot{\pi}^{\langle \mu\nu \rangle}$, we get only one independent 
equation as given above in eqn.(\ref{Bshear1}). 
It is important to note that the Kasner metric in the limit of 
$a=1, b=0, c=0$ reduces to that of the Minkowski metric in 
Milne coordinates. In this limit, the above second order evolution 
equations reduce to the evolution equations in the one dimensional 
Bjorken expansion case \cite{AJ}:
\begin{eqnarray}
\frac{d\epsilon}{d\tau} &= -\frac{1}{\tau}\left(\epsilon + P \right)  
+\frac{\pi}{\tau} \label{BED2} \\
\frac{d\pi}{d\tau} &= - \frac{\pi}{\tau_\pi} + \frac{4\beta_{\pi}}{3\tau} 
- \frac{38\pi}{21\tau}  \label{Bshear2}
\end{eqnarray}
%Nonzero third order terms are found to be
%\begin{align}
%\theta\equiv \nabla_\mu u^\mu = \frac{a+b+c}{\tau}
%\end{align}

Next, we compute the other terms appearing in the third order 
evolution equation of $\pi^{\mu\nu}$ and give the explicit expressions 
for the various terms. Components of $\pi_\gamma^{\langle \tau}\pi^{\tau 
\rangle\gamma}\theta$ are given by, 

\begin{eqnarray}{\label{seventhterm}}
\pi_\gamma^{\langle \tau}\pi^{\tau \rangle\gamma}\theta=0,
\pi_\gamma^{\langle x_1}\pi^{x_1 \rangle\gamma}\theta=
\frac{\pi^{2}(a+b+c)}{2\tau^{2a+1}},\nonumber\\
\pi_\gamma^{\langle x_2}\pi^{x_2 \rangle\gamma}\theta=
\frac{\pi^{2}(a+b+c)}{4\tau^{2b+1}}\nonumber\\
\pi_\gamma^{\langle x_3}\pi^{x_3 \rangle\gamma}\theta=
\frac{\pi^{2}(a+b+c)}{4\tau^{2c+1}}
\end{eqnarray}

Components of $\pi^{\mu\nu}\pi^{\rho\gamma}\sigma_{\rho\gamma}$ are : 
\begin{eqnarray}{\label{eightterm}}
\pi^{\tau\tau}\pi^{\rho\gamma}\sigma_{\rho\gamma}=0,\nonumber\\
\pi^{x_1x_1}\pi^{\rho\gamma}\sigma_{\rho\gamma}=
\frac{-\pi^{2}(2a-b-c)}{2\tau^{2a+1}},\nonumber\\
\pi^{x_2x_2}\pi^{\rho\gamma}\sigma_{\rho\gamma}=
\frac{\pi^{2}(2a-b-c)}{4\tau^{2b+1}}\nonumber\\
\pi^{x_3x_3}\pi^{\rho\gamma}\sigma_{\rho\gamma}=
\frac{\pi^{2}(2a-b-c)}{4\tau^{2c+1}}
\end{eqnarray}

Components of $\pi^{\rho\langle \mu}\pi^{\nu \rangle\gamma}\sigma_{\rho\gamma}$
are given by, 
\begin{eqnarray}{\label{ninethterm}}
\pi^{\rho\langle \tau}\pi^{\tau \rangle\gamma}\sigma_{\rho\gamma}=0,\nonumber\\
\pi^{\rho\langle x_1}\pi^{x_1\rangle\gamma}\sigma_{\rho\gamma}=
\frac{\pi^{2}(-2a+b+c)}{4\tau^{2a+1}},\nonumber\\
\pi^{\rho\langle x_2}\pi^{x_2 \rangle\gamma}\sigma_{\rho\gamma}=
\frac{\pi^{2}(a-b)}{4\tau^{2b+1}}\nonumber\\
\pi^{\rho\langle x_3}\pi^{x_3 \rangle\gamma}\sigma_{\rho\gamma}=
\frac{\pi^{2}(a-c)}{4\tau^{2c+1}}
\end{eqnarray}

For $\nabla^{\langle \mu}\left(\nabla_\gamma\pi^{\nu \rangle\gamma}\right)$
we have, 
\begin{eqnarray}{\label{eleventhterm}}
\nabla^{\langle \tau}\left(\nabla_\gamma\pi^{\tau \rangle\gamma}\right)=0\nonumber\\
\nabla^{\langle x_1}\left(\nabla_\gamma\pi^{x_1 \rangle\gamma}\right) =\frac{\pi}{6\tau^{2a+2}}(8a^{2}+2b^{2}+2c^{2}
-4ab+2bc-4ac)\nonumber\\
\nabla^{\langle x_2}\left(\nabla_\gamma\pi^{x_2\rangle\gamma}\right) =
\frac{\pi}{6\tau^{2b+2}}(-4a^{2}-4b^{2}+2c^{2}
+5ab-bc-ac)\nonumber\\
\nabla^{\langle x_3}\left(\nabla_\gamma\pi^{x_3\rangle\gamma}\right) =
\frac{\pi}{6\tau^{2c+2}}(-4a^{2}+2b^{2}-4c^{2}
-ab-bc+5ac)
\end{eqnarray}

For $\nabla_{\gamma}\left(\nabla^{\langle \mu}\pi^{\nu \rangle\gamma}\right)$,
we get, 
\begin{eqnarray}{\label{twelveterm}}
\nabla_{\gamma}\left(\nabla^{\langle \tau}\pi^{\tau \rangle\gamma}\right)=0\nonumber\\
\nabla_{\gamma}\left(\nabla^{\langle x_1}\pi^{x_1\rangle\gamma}\right) =
\frac{\pi}{6\tau^{2a+2}}(8a^{2}+2b^{2}+2c^{2}
+5ab+2bc+5ac)\nonumber\\
\nabla_{\gamma}\left(\nabla^{\langle x_2}\pi^{x_2 \rangle\gamma}\right) =
\frac{\pi}{6\tau^{2b+2}}(-4a^{2}-4b^{2}+2c^{2}
-4ab-bc-ac)\nonumber\\
\nabla_{\gamma}\left(\nabla^{\langle x_3}\pi^{x_3 \rangle\gamma}\right) =
\frac{\pi}{6\tau^{2c+2}}(-4a^{2}+2b^{2}-4c^{2}
-ab-bc-4ac)
\end{eqnarray}

Components of $\pi^{\mu\nu}\theta^2$ are given by, 
\begin{eqnarray}{\label{eighteenterm}}
\pi^{\tau \tau}\theta^2=0,
\pi^{x_1x_1}\theta^2=\frac{-\pi}{\tau^{2a}}\left(\frac{a+b+c}{\tau}\right)^{2} \nonumber\\
\pi^{x_2x_2}\theta^2=\frac{\pi}{2\tau^{2b}}\left(\frac{a+b+c}{\tau}\right)^{2}
\nonumber\\
\pi^{x_3x_3}\theta^2=\frac{\pi}{2\tau^{2c}}\left(\frac{a+b+c}{\tau}\right)^{2}
\end{eqnarray}

For $\nabla_{\gamma}\left(\tau_\pi\nabla^{\gamma}\pi^{\langle \mu\nu \rangle}
\right)$, we obtain, 
\begin{eqnarray}{\label{fourteenterm}}
\nabla_{\gamma}\left(\tau_\pi\nabla^{\gamma}\pi^{\langle \tau\tau \rangle}
\right)=0,\nonumber\\
\nabla_{\gamma}\left(\tau_\pi\nabla^{\gamma}\pi^{\langle x_1x_1 \rangle}\right)
=\frac{\pi(4a^{2}+b^{2}+c^{2})}{3\tau^{2a+2}},\nonumber\\
\nabla_{\gamma}\left(\tau_\pi\nabla^{\gamma}\pi^{\langle x_2x_2 \rangle}\right)
=\frac{\pi(-2a^{2}-2b^{2}+c^{2})}{3\tau^{2b+2}}\nonumber\\
\nabla_{\gamma}\left(\tau_\pi\nabla^{\gamma}\pi^{\langle x_3x_3 \rangle}\right)
=\frac{\pi(-2a^{2}+b^{2}-2c^{2})}{3\tau^{2c+2}}
\end{eqnarray}

All these expressions can be simplified further by using Kasner 
conditions. Here we have 
$\omega^{\mu\nu} = \dot {u}^{\mu} = \nabla^{\mu} \tau_{\pi} = 0$.
The other terms in the evolution equations as given below become zero, namely, 
\begin{eqnarray}{\label{zero-terms}}
\pi_{\gamma}^{\langle\mu}\omega^{\nu\rangle\gamma} & =0 \\ 
\nonumber
\pi^{\rho\langle\mu}\omega^{\nu\rangle\gamma}\pi_{\rho\gamma} & =0 \\
\nonumber
\nabla^{\langle\mu}\left(\pi^{\nu\rangle\gamma}\dot u_\gamma\tau_\pi\right) 
&=0 \\
\nonumber
\nabla_{\gamma}\left(\tau_\pi\dot u^{\langle\mu}\pi^{\nu\rangle\gamma}\right)
& =0 \\
\nonumber
\nabla_{\gamma}\left(\tau_\pi\dot u^{\gamma}\pi^{\langle\mu\nu\rangle}\right)
&=0 \\
\nonumber
\tau_\pi\omega^{\rho\langle\mu}\omega^{\nu\rangle\gamma}\pi_{\rho\gamma}&=0 \\
\nonumber
\tau_\pi\pi^{\rho\langle\mu}\omega^{\nu\rangle\gamma}\omega_{\rho\gamma} &=0 \\
\nonumber
\tau_\pi\pi_\gamma^{\langle\mu}\omega^{\nu\rangle\gamma}\theta &=0 
%\nonumber
\end{eqnarray}

Substituting the expressions for the above nonzero terms, the third order 
evolution equations for $x_1x_1$, $x_2x_2$, $x_3x_3$ components of the 
shear stress tensor are obtained as,  
%($\tau\tau$ component becomes identically zero),
\begin{eqnarray}{\label{x-x term}}
\frac{d\pi}{d\tau} &=-\frac{\pi}{\tau_\pi}-2\beta_\pi\frac{(-2a+b+c)}{3\tau}+\frac{\pi(-38a-23b-23c)}{21\tau}
+\frac{\pi^{2}(-803a+34b+34c)}{1470\beta_{\pi}\tau}\nonumber\\
&+\frac{\pi^{2}(106a^{2}-11b^{2}-11c^{2}-13ab-13ac-76bc)}{420\beta_{\pi}\tau}  
\end{eqnarray}
\begin{eqnarray}{\label{y-y term}}
\frac{d\pi}{d\tau} &= - \frac{\pi}{\tau_\pi} + 4\beta_\pi\frac{(a-2b+c)}{3\tau}+\frac{\pi(-38a-38b-8c)}{21\tau}
+\frac{\pi^{2}(-803a+199b-131c)}{1470\beta_{\pi}\tau}\nonumber\\
&+\frac{\pi^{2}(53a^{2}+53b^{2}-64c^{2}+25ab-38ac-38bc)}{210\beta_{\pi}\tau}  
\end{eqnarray}
\begin{eqnarray}{\label{z-z term}}
\frac{d\pi}{d\tau} &= - \frac{\pi}{\tau_\pi} + 4\beta_\pi\frac{(a+b-2c)}{3\tau}+\frac{\pi(-38a-8b-38c)}{21\tau}
+\frac{\pi^{2}(-803a-131b+199c)}{1470\beta_{\pi}\tau}\nonumber\\
&+\frac{\pi^{2}(53a^{2}-64b^{2}+53c^{2}+25ac-38ab-38bc)}{210\beta_{\pi}\tau}  
\end{eqnarray}

These equations can be further simplified by using Kasner conditions.  
One can check that adding the equations for the $x_1x_1$, 
$x_2x_2$ and $x_3x_3$ components, 
one gets only one independent equation, which is eq.(\ref{x-x term}). 
The evolution equation for the energy density $\epsilon$ is obtained as, 
\begin{eqnarray}\label{epsilon}
%\dot\epsilon + (\frac{\epsilon + P}{\tau}) + \frac{\pi}{2\tau}
%(b + c -2a) = 0
\frac{d\epsilon}{d\tau} = -\frac{1}{\tau} (\epsilon + P){\tau} 
- \frac{\pi (b + c -2a)}{2\tau}
\end{eqnarray}

In the limit $a=1, b=0, c=0$ for the Kasner parameters, 
the evolution equation for the energy density reduces
to that of the one dimensional Bjorken expansion case \cite{AJ}, 
\begin{eqnarray}\label{epsilon-1}
\frac{d\epsilon}{d\tau} &= -\frac{1}{\tau} (\epsilon + P -\pi)
\end{eqnarray}
and the third order evolution equation for the shear stress tensor 
$\pi^{\mu\nu}$ reduces to 
that of the equation for the Bjorken's one dimensional 
expansion case, namely, 
\begin{eqnarray}
\frac{d\pi}{d\tau} &= - \frac{\pi}{\tau_\pi} + 
\frac{4 \beta_\pi}{3\tau} - 
\frac{38}{21}\frac{\pi}{\tau} - \frac{72}{245}\frac{\pi^2}{\beta_\pi\tau}. 
\end{eqnarray}
By comparing the above equation with the third order evolution equation 
%eq.(\ref{Bshear}) 
in the one dimensional expansion case \cite{AJ}
\begin{eqnarray}
\frac{d\pi}{d\tau} &= - \frac{\pi}{\tau_\pi} +
\frac{4 \beta_\pi}{3\tau} -
\lambda \frac{\pi}{\tau} - \chi \frac{\pi^2}{\beta_\pi\tau},
\end{eqnarray}
the transport coefficients are given by, 
\begin{equation}\label{BTC}
\tau_\pi = \frac{\eta}{\beta_\pi}, \quad \beta_\pi = \frac{4P}{5}, 
\quad \lambda = \frac{38}{21}, \quad \chi = \frac{72}{245}.
\end{equation}
which matches with the results in the one dimensional expansion case. 

Here we would like to point out that one could have 
introduced three 
independent fields $\pi_i (i=1, 2, 3)$ for the shear stress tensor 
$\pi^{\mu\nu}$ instead of a
single function $\pi$, as has been introduced in the above discussion 
to charataterize $\pi^{\mu\nu}$. We have
explicitly checked that by making a general ansatz
for the shear stress tensor $\pi^{\mu\nu} = diag (0, \pi_1 \tau^{-2a},
\pi_2 \tau^{-2b}, \pi_3 \tau^{-2c})$ by introducing three independent
functions $\pi_1, \pi_2, \pi_3$, we get three independent third order
evolution equations (also second order) for the components of the shear 
stress tensor 
involving $\pi_1, \pi_2$ and $\pi_3$. These are given by, 
\begin{eqnarray}
\frac{d\pi_{1}}{d\tau} &= - \frac{\pi_{1}}{\tau_\pi}+
2\beta_\pi\frac{(-2a+b+c)}{3\tau}+\frac{\pi_{1}(-124a-64b-64c)}{63\tau}
+\frac{\pi_{2}(-10a+20b-10c)}{63\tau}+
\frac{\pi_{3}(-10a-10b+20c)}{63\tau}\nonumber\\
&-\frac{1}{735\beta_{\pi}\tau}[\pi_{1}^{2}(-386a-52b-52c)
+\pi_{2}^{2}(45a+155b+45c)+\pi_{3}^{2}(45a+45b+155c)\nonumber\\
&+\pi_{1}\pi_{2}(38a-76b+38c)+\pi_{1}\pi_{3}(38a+38b-76c)]\nonumber\\
&+\frac{\tau_{\pi}}{630 \tau^{2}}[\pi_{1}(212a^{2}-100b^{2}-100c^{2}-68ab-68ac\nonumber\\
&-200bc)+\pi_{2}(-156b^{2}-84ab-48bc)
+\pi_{3}(-156c^{2}-84ac-48bc)]
\end{eqnarray}
\begin{eqnarray}
%{\label{x-x term}}
\frac{d\pi_{2}}{d\tau} &=-\frac{\pi_{2}}{\tau_\pi}+
2\beta_\pi\frac{(a-2b+c)}{3\tau}+\frac{\pi_{1}(20a-10b-10c)}{63\tau}
+\frac{\pi_{2}(-64a-124b-64c)}{63\tau}+\frac{\pi_{3}(-10a-10b+20c)}{63\tau}\nonumber\\
&-\frac{1}{735\beta_{\pi}\tau}[\pi_{1}^{2}(155a+45b+45c)
+\pi_{2}^{2}(-52a-386b-52c)+\pi_{3}^{2}(45a+45b+155c)\nonumber\\
&+\pi_{1}\pi_{2}(-76a+38b+38c)+\pi_{2}\pi_{3}(38a+38b-76c)]\nonumber\\
&+\frac{\tau_{\pi}}{315\tau^{2}}[\pi_{1}(-78a^{2}-42ab-24ac)+\pi_{2}(-50a^{2}+106b^{2}\nonumber\\
&-50c^{2}-34ab-34bc
-100ac)+\pi_{3}(-78c^{2}-24ac-42bc)]
\end{eqnarray}
\begin{eqnarray}
%{\label{y-y term}}
\frac{d\pi_{3}}{d\tau} &=-\frac{\pi_{3}}{\tau_\pi} +
2\beta_\pi\frac{(a+b-2c)}{3\tau}+\frac{\pi_{1}(20a-10b-10c)}{63\tau}
+\frac{\pi_{2}(-10a+20b-10c)}{63\tau}+\frac{\pi_{3}(-64a-64b-124c)}{63\tau}\nonumber\\
&-\frac{1}{735\beta_{\pi}\tau}[\pi_{1}^{2}(155a+45b+45c)
+\pi_{2}^{2}(45a+155b+45c)+\pi_{3}^{2}(-52a-52b-386c)\nonumber\\
&+\pi_{1}\pi_{3}(-76a+38b+38c)+\pi_{2}\pi_{3}(38a-76b+38c)]\nonumber\\
&+\frac{\tau_{\pi}}{315\tau^{2}}[\pi_{1}(-78a^{2}-24ab-42ac)
+\pi_{2}(-78b^{2}-24ab-42bc)+\pi_{3}(-50a^{2}-50b^{2}\nonumber\\
&+106c^{2}-34ac-34bc-100ab)]
\end{eqnarray}
Since $\pi^{\mu\nu}$ is traceless, for the Kasner metric $g_{\mu\nu}$, 
we get the condition
\begin{equation}
\pi_1 + \pi_2 + \pi_3 = 0
\end{equation}
Using this tracelessness condition, we find
that adding the equations for $\pi_2$ and $\pi_3$, we get precisely the
equation for $\pi_1$ (where, $\dot\pi^{\langle 1 1 \rangle} = 
\frac{-1}{\tau^{2 a}} \frac{d\pi_1} {d\tau}$). Below we show this 
explicitly. For making this check, 
we have also computed all the relevant terms appearing in the third order
evolution equations in term of $\pi_1, \pi_2, \pi_3$ (we give these 
expressions in the appendix).  
Adding the equations for $\frac{d\pi_{2}}{d\tau}$ and $\frac{d\pi_{3}}{d\tau}$,
and using $\pi_2 + \pi_3 = -\pi_1$, we obtain,
\begin{eqnarray}{\label{11-final-term}}
\frac{d\pi_{1}}{d\tau} &= - \frac{\pi_{1}}{\tau_\pi}+ 2\beta_\pi\frac{(-2a+b+c)}{3\tau}+\frac{\pi_{1}(-40a+20b+20c)}{63\tau}
+\frac{\pi_{2}(74a+104b+74c)}{63\tau}+\frac{\pi_{3}(74a+74b+104c)}{63\tau}\nonumber\\
&-\frac{1}{735\beta_{\pi}\tau}[\pi_{1}^{2}(-310a-90b-90c)
+\pi_{2}^{2}(7a+231b+7c)+\pi_{3}^{2}(7a+7b+231c)\nonumber\\
&+\pi_{1}\pi_{2}(76a-38b-38c)+\pi_{1}\pi_{3}(76a-38b-38c)\nonumber\\
&+\pi_{2}\pi_{3}(-76a+38b+38c)]
+\frac{\tau_{\pi}}{315 \tau^{2}} [\pi_{1}(156 a^{2} +
66 ab + 66ac) + \pi_2 (50 a^2 -28 b^2 \nonumber\\
&+ 50c^2+58ab + 76 bc + 100 ac)+\pi_3 (50 a^2\nonumber\\
& + 50 b^2 -28 c^2 + 76 bc + 58 ac + 100ab)]
\end{eqnarray}
This can be simplified to,
\begin{eqnarray}{\label{eta-eta term}}
\frac{d\pi_{1}}{d\tau} &= - \frac{\pi_{1}}{\tau_\pi}+ 2\beta_\pi\frac{(-2a+b+c)}{3\tau}+\frac{\pi_{1}(-124a-64b-64c)}{63\tau}
+\frac{\pi_{2}(-10a+20b-10c)}{63\tau}
+\frac{\pi_{3}(-10a-10b+20c)}{63\tau}\nonumber\\
%&+\frac{(\pi_1+\pi_2+\pi_3)(84a+84b+84c)}{63\tau} \nonumber\\
&-\frac{1}{735\beta_{\pi}\tau}[\pi_{1}^{2}(-386a-52b-52c)
+\pi_{2}^{2}(45a+155b+45c)+\pi_{3}^{2}(45a+45b+155c)\nonumber\\
&+\pi_{1}\pi_{2}(38a-76b+38c)+\pi_{1}\pi_{3}(38a+38b-76c)]\nonumber\\
%       &+ \frac{1}{735\beta_{\pi}\tau}((-38a-38b-38c)
%(\pi_1+\pi_2+\pi_3)^2 + 114 a \pi_1 (\pi_1+\pi_2+\pi_3) \nonumber\\
%&+114 b \pi_2 (\pi_1+\pi_2+\pi_3)+114 c \pi_3 (\pi_1+\pi_2+\pi_3))
%\nonumber\\
&+\frac{\tau_{\pi}}{630 \tau^{2}}[\pi_{1}(212a^{2}-100b^{2}-100c^{2}-68ab-68ac-200bc)+\pi_{2}(-156b^{2}
\nonumber\\
&-84ab-48bc)+\pi_{3}(-156c^{2}
-84ac-48bc)]+\frac{(\pi_{1}+\pi_{2}+\pi_{3})(84a+84b+84c)}{63\tau} \nonumber\\
&+ \frac{1}{735\beta_{\pi}\tau}[(-38a-38b-38c)
(\pi_1+\pi_2+\pi_3)^2 
+ 114 (a \pi_1 + b \pi_{2} + c \pi_{3})(\pi_{1}+\pi_{2}+\pi_{3})] \nonumber\\
&+\frac{\tau_{\pi}}{630 \tau^{2}}
(\pi_{1}+\pi_{2}+\pi_{3})(100a^2 + 100b^2+100c^2
+200ab+200bc+200ac)
\end{eqnarray}
where the last three terms are zero as $\pi_1 + \pi_2 + \pi_3 = 0$.
Rest of
the terms are precisely the expression for the $\frac{d\pi_1}{d\tau}$
equation. Hence there is no inconsistency and there is only one
independent ordinary differential equation (ODE).
Hence, for simplicity, we have made the assumption that
$\pi_{\mu\nu}$
could be characterized by a single function $\pi$ and the three ODEs
involving $x_1x_1, x_2x_2, x_3x_3$ components of the shear tensor 
are not totally
independent, rather adding the $x_2x_2$ and $x_3x_3$ equations ,
it reduces to the equation for the $x_1x_1$ component. This we have 
already discussed with reference to equations (40), (41) and (42) 
in this section. We have shown above 
that this is also the case for the general ansatz in terms
of $\pi_1, \pi_2$ and $\pi_3$. We have also checked that putting
$\pi_1 = - \pi, \pi_2 = \frac{\pi}{2}, \pi_3 = \frac{\pi}{2}$ in the
ODEs, we get back the earlier equations in terms of $\pi$, namely, 
equations (40), (41) and (42) (where we have used the first order 
relation $\pi = \frac{4\beta_{\pi}\tau_{\pi}}{3 \tau}$ for rewriting 
the last term in terms of the coefficient $\frac{\pi^2}{\beta_{\pi}\tau}$. 
All these expressions also reduce to the
Bjorken case for $a=1, b=0=c$.
Hence the ansatz is consistent. This is also true for the second
order evolution equation.  

%%%%%%%%%%%%%%%%%%%%%%%%%%%%%%%%%%%%%%%%%%%%%%%%%%%%%%%%%%%%%%%%%%%%%%%%%%

\section{Summary and discussion}
\label{sec:Summary}
\setcounter{equation}{0}

In this work, we have studied relativistic viscous hydrodynamics 
from kinetic theory to second and third order in gradient expansion. 
We have considered Kasner space-time as the LRF
of the three dimensional anisotropic expansion of a conformal fluid 
and have obtained the second and 
third order evolution equations for the shear stress tensor and energy 
density. 
We have used the iterative solutions of the Boltzmann 
equation for the nonequilibrium distribution function in RTA. 
We have shown that our 
results for the three dimensional expansion agree with the one dimensional 
Bjorken expansion case in appropriate limit of the anisotropy parameters. 
%%%%%%%%%%%%%%%%%%%%%%%%%%%%%%%%%%%%%%%%%%%%%%%%%%%%%%
For a general 
fluid in Kasner space-time, the system may not be conformal, in which 
case the stress energy tensor will not be traceless. It is a challenging 
topic to study the evolution of the shear stress tensor for a nonconformal 
system with nonzero bulk viscosity in a general curved background. 
Nonconformal fluid models contain a much larger number of transport 
coefficients. There has been some progress in related aspects from various 
approaches (see for example, 
\cite{Romatschke:2009, Denicol:2014, Finazzo:2014}).
It will be certainly interesting to explore the dissipative evolution 
equations in a higher order gradient expansion involving nonconformal 
fluid in Kasner space-time.
%%%%%%%%%%%%%%%%%%%%%%%%%%%%%%%%%%%%%%%%%%%%%%%%%%%%%%
One can also work out the higher order corrections to entropy four current 
in our setup of anisotropic expansion. It will be interesting to 
relate our results for anisotropic 
space-time to anisotropic hydrodynamics 
(see ref.\cite{Strickland:2014} for a review on anisotropic hydrodynamics). 
It will be worth exploring higher order dissipative hydrodynamics for 
Gubser flow \cite{Chandrodoy:2018} and relate it to anisotropic hydrodynamics 
in the present scenario. 
In the context of anisotropic expansion of the fluid, it will also be
interesting to study the higher order corrections in relativistic 
hydrodynamics in extended relaxation time approximation 
\cite{Rocha-etal:2021, Jaiswal-etal:2021}, where the 
relaxation time depends on particle energy. We hope to report 
on these issues in future. 

\vskip 10pt
\noindent
{\bf Acknowledgements} \\
We would like to thank Amaresh Jaiswal for helpful discussions.
\vskip 10pt
%%%%%%%%%%%%%%%%%%%%%%%%%%%%%%%%%%%%%%%%%%%%%%%%%%%%%%%%%%%%%%%%

\section{Appendix}
\label{sec:Appendix}
\setcounter{equation}{0}
We assume that the shear stress tensor is diagonal and 
is characterized 
by three functions $\pi_1, \pi_2$ and $\pi_3$, namely,  
$\pi^{\mu\nu}\equiv\mathrm{diag}(0,\pi_{1}\tau^{-2a},
\pi_{2}{\tau^{-2b}},\pi_{3}{\tau^{-2c}})$.
The components of $\dot{\pi}^{\langle \mu\nu \rangle}$ are 
given by, 
\begin{eqnarray}{\label{zeroterm}}
\dot{\pi}^{\langle \tau\tau \rangle}= 0, \,
\dot{\pi}^{\langle x_1x_1 \rangle}=\frac{1}{\tau^{2a}}\frac{d\pi_1}{d\tau}, \,
\nonumber\\
\dot{\pi}^{\langle x_2x_2 \rangle}=\frac{1}{\tau^{2b}}\frac{d\pi_2}{d\tau}, \,
\dot{\pi}^{\langle x_3x_3 \rangle}=\frac{1}{\tau^{2c}}\frac{d\pi_3}{d\tau}
\end{eqnarray}

Components of other terms appearing in the third order 
evolution equation are obtained as,
\begin{eqnarray}{\label{fourthterm}}
\pi_\gamma^{\langle \tau}\sigma^{\tau\rangle\gamma}= 0\nonumber\\
\pi_\gamma^{\langle x_{1}}\sigma^{x_{1}\rangle\gamma}=\frac{1}
{9 \tau^{2a+1}}(-2\pi_{1}(-2a+b+c)\nonumber\\
+\pi_{2}(a-2b+c)+\pi_{3}(a+b-2c)\nonumber\\
\pi_\gamma^{\langle x_{2}}\sigma^{x_{2} \rangle\gamma}=
\frac{1}{9 \tau^{2b+1}}(\pi_{1}(-2a+b+c)\nonumber\\
-2\pi_{2}(a-2b+c)+\pi_{3}(a+b-2c)\nonumber\\
\pi_\gamma^{\langle x_{3}}\sigma^{x_{3} \rangle\gamma}=\frac{1}
{9 \tau^{2c+1}}(\pi_{1}(-2a+b+c)\nonumber\\
+\pi_{2}(a-2b+c)-2\pi_{3}(a+b-2c)
\end{eqnarray}

Components of $\pi_\gamma^{\langle \tau}\pi^{\tau 
\rangle\gamma}\theta$ are given by, 
\begin{eqnarray}{\label{seventhterm}}
\pi_\gamma^{\langle \tau}\pi^{\tau \rangle\gamma}\theta=0\nonumber\\
\pi_\gamma^{\langle x_{1}}\pi^{x_{1} \rangle\gamma}\theta=
\frac{(-2\pi_{1}^{2}+\pi_{2}^{2}+\pi_{3}^{2})(a+b+c)}{3\tau^{2a+1}}\nonumber\\
\pi_\gamma^{\langle x_{2}}\pi^{x_{2} \rangle\gamma}\theta=\frac{(\pi_{1}^{2}-
2\pi_{2}^{2}+\pi_{3}^{2})(a+b+c)}{3\tau^{2b+1}}\nonumber\\
\pi_\gamma^{\langle x_{3}}\pi^{x_{3} \rangle\gamma}\theta=\frac{(\pi_{1}^{2}+
\pi_{2}^{2}-2\pi_{3}^{2})(a+b+c)}{3\tau^{2c+1}}
\end{eqnarray}

Components of $\pi^{\mu\nu}\pi^{\rho\gamma}
\sigma_{\rho\gamma}$ are : 
\begin{eqnarray}{\label{eightterm}}
\pi^{\tau\tau}\pi^{\rho\gamma}\sigma_{\rho\gamma}=0\nonumber\\
\pi^{x_{1}x_{1}}\pi^{\rho\gamma}\sigma_{\rho\gamma}=\frac{\pi_{1}}{3\tau^{2a+1}}[(\pi_{1}(-2a+b+c)\nonumber\\
+\pi_{2}(a-2b+c)+\pi_{3}(a+b-2c)]\nonumber\\
\pi^{x_{2}x_{2}}\pi^{\rho\gamma}\sigma_{\rho\gamma}=\frac{\pi_{2}}{3\tau^{2b+1}}[(\pi_{1}(-2a+b+c)\nonumber\\
+\pi_{2}(a-2b+c)+\pi_{3}(a+b-2c)]\nonumber\\
\pi^{x_{3}x_{3}}\pi^{\rho\gamma}\sigma_{\rho\gamma}=\frac{\pi_{3}}{3\tau^{2c+1}}[(\pi_{1}(-2a+b+c)\nonumber\\
+\pi_{2}(a-2b+c)+\pi_{3}(a+b-2c)]
\end{eqnarray}

Components of $\pi^{\rho\langle \mu}\pi^{\nu \rangle\gamma}
\sigma_{\rho\gamma}$ are given by, 
\begin{eqnarray}{\label{ninethterm}}
\pi^{\rho\langle \tau}\pi^{\tau \rangle\gamma}\sigma_{\rho\gamma}=0\nonumber\\
\pi^{\rho\langle x_{1}}\pi^{x_{1}\rangle\gamma}\sigma_{\rho\gamma}=
\frac{1}{9\tau^{2a+1}}[(2\pi_{1}^{2}(-2a+b+c)\nonumber\\
-\pi_{2}^{2}(a-2b+c)-\pi_{3}^{2}(a+b-2c)]\nonumber\\
\pi^{\rho\langle x_{2}}\pi^{x_{2} \rangle\gamma}\sigma_{\rho\gamma}=
\frac{1}{9\tau^{2b+1}}[-\pi_{1}^{2}(-2a+b+c)\nonumber\\
+2\pi_{2}^{2}(a-2b+c)-\pi_{3}^{2}(a+b-2c)]\nonumber\\
\pi^{\rho\langle x_{3}}\pi^{x_{3} \rangle\gamma}\sigma_{\rho\gamma}=
\frac{1}{9\tau^{2c+1}}[-\pi_{1}^{2}(-2a+b+c)\nonumber\\
-\pi_{2}^{2}(a-2b+c)+2\pi_{3}^{2}(a+b-2c)]
\end{eqnarray}

For $\nabla^{\langle \mu}\left(\nabla_\gamma\pi^{\nu 
\rangle\gamma}\right)$ we have, 
\begin{eqnarray}{\label{eleventhterm}}
\nabla^{\langle \tau}\left(\nabla_\gamma\pi^{\tau \rangle\gamma}\right)=0
\nonumber\\
\nabla^{\langle x_{1}}\left(\nabla_\gamma\pi^{x_{1} \rangle\gamma}\right)=
\frac{1}{3\tau^{2a+2}}(\pi_{1}(-4a^{2}+ab+ac)\nonumber\\
+\pi_{2}(2b^{2}-2ab+bc)+\pi_{3}(2c^{2}+bc-2ac))\nonumber\\
\nabla^{\langle x_{2}}\left(\nabla_\gamma\pi^{x_{2} \rangle\gamma}\right)=
\frac{1}{3\tau^{2b+2}}(\pi_{1}(2a^{2}-2ab+ac)\nonumber\\+\pi_{2}(-4b^{2}+ab+bc)+\pi_{3}(2c^{2}-2bc+ac))\nonumber\\
\nabla^{\langle x_{3}}\left(\nabla_\gamma\pi^{x_{3} \rangle\gamma}\right)=
\frac{1}{3\tau^{2c+2}}(\pi_{1}(2a^{2}+ab-2ac)\nonumber\\+\pi_{2}(2b^{2}+ab-2bc)
+\pi_{3}(-4c^{2}+bc+ac))
\end{eqnarray}

For $\nabla_{\gamma}\left(\nabla^{\langle \mu}\pi^{\nu 
\rangle\gamma}\right)$, we get, 
\begin{eqnarray}{\label{twelveterm}}
\nabla_{\gamma}\left(\nabla^{\langle \tau}\pi^{\tau \rangle\gamma}\right)=0
\nonumber\\
\nabla_{\gamma}\left(\nabla^{\langle x_{1}}\pi^{x_{1} \rangle\gamma}\right)=
\frac{1}{3\tau^{2a+2}}(\pi_{1}(-4a^{2}-2ab-2ac)\nonumber\\
+ \pi_{2}(2b^{2}+ab+bc)+\pi_{3}(2c^{2}+bc+ac))\nonumber\\
\nabla_{\gamma}\left(\nabla^{\langle x_{2}}\pi^{x_{2} \rangle\gamma}\right)=
\frac{1}{3\tau^{2b+2}}(\pi_{1}(2a^{2}+ab+ac)\nonumber\\
+\pi_{2}(-4b^{2}-2ab-2bc)+\pi_{3}(2c^{2}+bc+ac))\nonumber\\
\nabla_{\gamma}\left(\nabla^{\langle x_{3}}\pi^{x_{3} \rangle\gamma}\right)=
\frac{1}{3\tau^{2c+2}}(\pi_{1}(2a^{2}+ab+ac)\nonumber\\
+\pi_{2}(2b^{2}+ab+bc)+\pi_{3}(-4c^{2}-2bc-2ac))
\end{eqnarray}

For $\nabla_{\gamma}\left(\tau_\pi\nabla^{\gamma}
\pi^{\langle \mu\nu \rangle} \right)$, we obtain, 
\begin{eqnarray}{\label{fourteenterm}}
\nabla_{\gamma}\left(\tau_\pi\nabla^{\gamma}\pi^{\langle \tau\tau \rangle}
\right)=0\nonumber\\
\nabla_{\gamma}\left(\tau_\pi\nabla^{\gamma}\pi^{\langle x_{1}x_{1} \rangle}
\right)=\frac{2\tau_\pi(-2a^{2}\pi_{1}+b^{2}\pi_{2}+c^{2}\pi_{3})}{3\tau^{2a+2}}\nonumber\\
\nabla_{\gamma}\left(\tau_\pi\nabla^{\gamma}\pi^{\langle x_{2}x_{2} \rangle}
\right)=\frac{2\tau_\pi(a^{2}\pi_{1}-2b^{2}\pi_{2}+c^{2}\pi_{3})}{3\tau^{2b+2}}
\nonumber\\
\nabla_{\gamma}\left(\tau_\pi\nabla^{\gamma}\pi^{\langle x_{3}x_{3} \rangle}
\right)=\frac{2\tau_\pi(a^{2}\pi_{1}+b^{2}\pi_{2}-2c^{2}\pi_{3})}{3\tau^{2c+2}}
\end{eqnarray}

Components of $\pi^{\mu\nu}\theta^2$ are given by, 
\begin{eqnarray}{\label{eighteenterm}}
\pi^{\tau \tau}\theta^2=0\nonumber\\
\pi^{x_{1}x_{1}}\theta^2=\frac{\pi_{1}}{\tau^{2a}}\left(\frac{a+b+c}{\tau}
\right)^{2}\nonumber\\
\pi^{x_{2}x_{2}}\theta^2=\frac{\pi_{2}}{\tau^{2b}}\left(\frac{a+b+c}{\tau}
\right)^{2}\nonumber\\
\pi^{x_{3}x_{3}}\theta^2=\frac{\pi_{3}}{\tau^{2c}}\left(\frac{a+b+c}{\tau}
\right)^{2}
\end{eqnarray}
These expressions have beeen used to obtain the differential equations 
for the components of $\pi^{\mu\nu}$ in terms of 
$\pi_1, \pi_2, \pi_3$. All these expressions reduce to our earlier 
results where we have considered a single function $\pi$ to characterize
the shear stress tensor $\pi^{\mu\nu}$.  

%%%%%%%%%%%%%%%%%%%%%%%%%%%%%%%%%%%%%%%%%%%%%%%%%%%%%%%%%%%%%%%%
%%%%%%%%%%%%%%%%%%%%%%%%%%%%%%%%%%%%%%%%%%%%%%%%%%%%%%%%%%%%%%%%
\providecommand{\href}[2]{#2}

%%%%%%%%%%%%%%%%%%%%%%%%%%%%%%%%%%%%%%%%%%%%%%%%%%%%%%%%%%%%%%%%%%%%%
\end{document}